\documentclass[12pt]{article}
\usepackage{epsf}
\renewcommand{\arraystretch}{1.5}
\input prepictex
\input pictex
\input postpictex
\newdimen\tdim
\tdim=\unitlength
\def\stpltsmbl{\setplotsymbol ({\small .})}

\def\tarrow{\arrow <5\tdim> [.3,.6]}

\def\tr{\mathord{{\rm tr}}}
\def\d{\mathord{{\rm d}}}
\def\D{\mathord{{\rm D}}}
\def\Pf{\mathord{{\rm Pf}}}

\def\cD{{\cal D}}

\begin{document}

\parindent = 12pt

\pagestyle{plain}
\pagenumbering{arabic}
\newcommand{\Lam}[1]{\Lambda_#1^{4}}
\newcommand{\tLam}[1]{\tilde{\Lambda}_#1^{4}}
\newcommand{\preprint}[1]{\rule{0pt}{8pt} \scriptsize #1}

\title{Quantum Modified Mooses\thanks{Research supported in part by the
National
Science Foundation under grant number NSF-PHY/98-02709.}}
\author{Spencer Chang\thanks{chang@physics.harvard.edu}~
and Howard Georgi\thanks{georgi@physics.harvard.edu}\\
  \small\sl Lyman Laboratory of Physics \\
  \small\sl Harvard University \\
  \small\sl Cambridge, MA 02138}
\date{}

  \maketitle
  \begin{picture}(0,0)
    \put(400,200){\shortstack{
        \preprint HUTP-02/A044\\
	\preprint hep-th/0209038\\
        \rule{0pt}{8pt} }}
  \end{picture}

\begin{abstract}

We summarize our findings on the quantum moduli constraints and superpotentials
of an infinite family of moose extensions of $n_f = n_c$ SUSY QCD.
For $n_c=2$, we perform concrete calculations using traditional integrating out
techniques as well as
Intriligator's ``integrating in'' technique.
Checking the constraints and superpotentials in the limits of setting
$\Lambda$'s to zero or integrating out mass terms, we find that the quantum
moduli constraints are local in theory space and are equivalent to
a consistent structure of ``splitting relations'' amongst the different
theories. Extending the results to arbitrary $n_c$, we show that the
splitting relations, along with a set of rules for flowing from a high
energy theory to a low energy theory, incorporate much of the physics of the
moose chain. The relations
can be used both to simplify perturbative calculations of symmetry breaking
and to incorporate nonperturbative effects. 

\end{abstract}
  \thispagestyle{empty} \setcounter{page}{0}

\newpage
\section{Introduction}
\noindent

In the past year, the method of ``deconstruction'' \cite{Arkani-Hamed:2001ca, Hill:2000mu} has
taken hold as a new way of model building and addressing unresolved
problems in particle physics.  It was first presented as a way of
dynamically generating extra dimensions in certain energy regimes.  Even far from the
continuum limit, extra-dimensional properties (such as locality) persist in the
``theory'' space of these models.  This has provided a new tool to
address unresolved issues in novel ways and also by ``deconstruction''
of previously known extra-dimensional mechanisms \cite{Csaki:2002fy}.

\begin{figure}[htb]
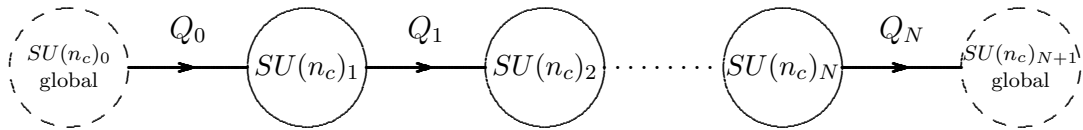

\begin{center}
$$\beginpicture
\setcoordinatesystem units <.9\tdim,.9\tdim>
\circulararc 360 degrees from 25 0 center at 0 0
\circulararc 360 degrees from 100 25 center at 100 0
\circulararc 360 degrees from 200 25 center at 200 0
\setdashes
\circulararc 360 degrees from -100 25 center at -100 0
\circulararc 360 degrees from 300 25 center at 300 0
\setsolid
\put {$Q_0$} [b] at -50 10
\put {$Q_1$} [b] at 50 10
\put {$Q_N$} [b] at 250 10
\put {\small$SU(n_c)_1$} at 0 0
\put {\small$SU(n_c)_2$} at 100 0
\put {\small$SU(n_c)_N$} at 200 0
\put {\scriptsize\stack{$SU(n_c)_{N+1}$,global}} at 300 0
\put {\scriptsize\stack{$SU(n_c)_{0}$,global}} at -100 0
\stpltsmbl
\tarrow from -75 0 to -48 0
\plot -48 0 -25 0 /
\tarrow from 25 0 to 52 0
\plot 52 0 75 0 /
\tarrow from 225 0 to 252 0
\plot 252 0 275 0 /
\setdots
\plot 125 0 175 0 /
\linethickness=0pt
\putrule from -125 0 to 325 0
\endpicture$$
\caption{Chain Extension of SUSY QCD with $n_f = n_c$}
\label{fig1}
\end{center}
\end{figure}

Crucial to the ``deconstruction'' mechanism is the structure of the allowed
vacua of the theory.
It is therefore interesting to
investigate models where the vacuum structure can be analyzed
nonperturbatively.  A particularly
simple family of models is SUSY $SU(n_c)^N$ gauge theories, which in
a moose diagram appears to be a discretization of an extra 5th dimension
(see fig.~\ref{fig1}).

In this moose diagram, each solid circle represents an $SU(n_c)$ gauge group,
each dashed circle a global $SU(n_c)$ group, and each $Q$ a chiral superfield.
Now, according to moose conventions \cite{Georgi:1985hf}, the field $Q_i$
transforms as a $(n_c,\bar n_c)$ under $(SU(n_c)_i,SU(n_c)_{i+1})$.  As one
can see this
theory is free of gauge anomalies and is a natural moose extension of $n_f
= n_c$ SUSY QCD.  Now if $<Q_i> = aI$ for $i = 1, \cdots
N-1$, a perturbative analysis suggests that this theory deconstructs 5-d $n_f =
n_c$ SUSY QCD compactified on a $S_1/Z_2$ orbifold.  We wish to see
if this feature is maintained nonperturbatively, and discovering the
nonperturbative vacua for these models will be the main topic of this
paper.  Thus, in essence, we will attempt to extend Seiberg's
nonperturbative SUSY QCD results \cite{Intriligator:1995au} to
this interesting group of models. We will find that the results are simple,
beautiful and useful. We will describe a simple set of ``splitting
functions'' that incorporate the structure of the moduli space, and thus
also much of the physics. These splitting functions can be used both to
simplify perturbative calculations and to incorporate nonperturbative
effects.

The layout of this paper is as follows: Sections 2-7 focus on $n_c=2$ ---
in Section 2 we give a quick review
of the standard lore for SUSY QCD required for our analysis; in Section 3
we discuss some preliminaries; in Section 4 we look at simple example
calculations and describe how an iterative process determines the results
for all N; in Section 5 we discuss the computational results and describe the
general splitting functions (and explain why we call them ``splitting
functions''); in Section 6 we demonstrate all the possible
consistency checks we can perform; in Section 7, we discuss spontaneous
symmetry breaking, emphasizing that our splitting relations are local in
theory space; in Section 8, we generalize our results to
$SU(n_c)^N$; in Section 9 we try to analyze power-law running; and in Section 10 we
conclude and look at questions that should be addressed by
further research.  In addition, in the appendix we discuss an argument that
completes the analysis of the constraints for $SU(2)$.

\section{Standard Lore}
For the purposes of this paper, the results from Seiberg,
{\it et al.\/} \cite{Intriligator:1995au} that we are most interested in are those for SUSY QCD
with $n_f \leq n_c$.  Using arguments based on holomorphy and symmetry, they deduced
that at the nonperturbative level, any generated superpotential had the
form

\begin{equation}
W = c_{n_f,n_c}\left(\frac{\Lambda^{3n_c-n_f}}{\det
(Q\tilde{Q})}\right)^{\frac{1}{n_c-n_f}}.
\label{eq:ADS}
\end{equation}
To determine if this is actually generated, they argued that for $n_f = n_c
- 1$ this can be reliably calculated via an instanton calculation (since
the gauge symmetry is entirely broken by the vev).  Finnell and Pouliot's
calculation in $SU(2)$ determines that $c_{1,2} = 1$ in $\overline{DR}$
\cite{Finnell:1995dr}.  Integrating out quarks gives recurrence relations for the
$c$'s that fixes $c_{n_f,n_c} = n_c-n_f$.  Two things to note about this
superpotential are that it doesn't make sense for $n_f \geq n_c$ (since
either $\det (Q\tilde{Q}) = 0$ or the power diverges) and that for $n_f <
n_c$ the vacuum is pushed off to infinity.

In the limit $n_f \to 0$, (\ref{eq:ADS}) gives the gaugino condensation
superpotential
\begin{equation}
W_{gaugino} = n_c\Lambda^3.
\end{equation}
Notice that there are $n_c$ inequivalent vacua
(corresponding to a $\theta \to \theta + 2\pi$ transformation where
$\Lambda^3 \to
e^{2\pi i/n_c}\Lambda^3$) as expected by the Witten index analysis.

Finally, they discovered that at $n_f = n_c = 2$, the classical moduli
constraint $\Pf (M) = 0$ is modified.  Incorporating the nonperturbative
effects, they found that the full Quantum Modified Moduli Space (QMMS)
constraint was
\begin{equation}
\Pf (M) = \Lambda^{4}.
\label{eq:plussign}
\end{equation}
For one site with $n_c\neq2$ the QMMS condition in our notation should be
written as
\begin{equation}
\det(Q_0Q_1)=\det Q_0 \det Q_1-\Lambda_1^{2n_c}.
\label{qmmsnc}
\end{equation}

\section{Preliminaries and a teaser\label{teaser}}
Consider the chain model we introduced in fig.~\ref{fig1}. Just as in
$n_f = n_c$ SUSY QCD, there is an anomaly free $U(1)_R$ symmetry under
which each of the $Q$'s are uncharged.  Assuming that this $U(1)_R$ holds
nonperturbatively, there can be no superpotential generated.  Thus the
moduli space is not lifted, and a quantum moduli space should exist.
According to Luty and Taylor \cite{Luty:1995sd}, the moduli space can be
parameterized by a set of gauge invariant monomials.  In the chain model,
this set comprises
\begin{equation}
\begin{array}{c}
\det (Q_0), \: \det (Q_1), \: \cdots, \: \det (Q_N)\\
\det (Q_0Q_1), \: \det (Q_1Q_2), \: \cdots, \: \det (Q_{N-1}Q_N)\\
\cdots\\
\det (Q_0Q_1\cdots Q_{N-1}), \: \det (Q_1Q_2\cdots Q_N)\\
Q_0Q_1\cdots Q _N\,.
\end{array}
\end{equation}
This gives us $n_c^2 + \frac{N(N+3)}{2}$ moduli fields.  A general vacuum
breaks all of the gauge groups, which eats up $N(n_c^2-1)$ of the scalars in
the $Q$'s.  This leaves $(N+1)n_c^2 - N(n_c^2-1) = N+n_c^2$ independent moduli
fields.  Thus we expect there to be $n_c^2 + \frac{N(N+3)}{2} - (N+n_c^2) =
\frac{N(N+1)}{2}$ constraints amongst the moduli fields above.
Classically, these constraints are just given by

\begin{equation}
\det (Q_iQ_{i+1}\cdots Q_j) = \det (Q_i)\det (Q_{i+1})\cdots \det (Q_j).
\end{equation}

A standard procedure when analyzing these theories is to isolate one's
attention to an independent set of gauge invariant operators.  However, in
doing so, one throws out a significant amount of information, specifically
the very constraints that we are trying to analyze!  On the other hand,
keeping all gauge invariant operators can make the problem intractable
since manipulating the operators requires a knowledge of their
interdependence.

With this in mind, in the next section, we choose a set of gauge invariant
operators that is
complete, but with just one expected constraint amongst them.  The
particular set of operators is
\begin{equation}
\det (Q_0), \: \det (Q_1), \: \cdots, \: \det (Q_N), \: Q_0\cdots Q_N.
\end{equation}
In the later sections, when we discover the constraint that relates these
operators, we will refer to it as the ``highest'' constraint.

\begin{figure}[htb]
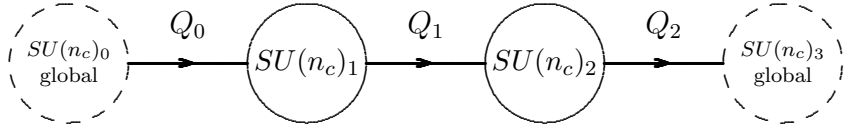

\begin{center}
$$\beginpicture
\setcoordinatesystem units <.9\tdim,.9\tdim>
\circulararc 360 degrees from 25 0 center at 0 0
\circulararc 360 degrees from 100 25 center at 100 0
\setdashes
\circulararc 360 degrees from -100 25 center at -100 0
\circulararc 360 degrees from 200 25 center at 200 0
\setsolid
\put {$Q_0$} [b] at -50 10
\put {$Q_1$} [b] at 50 10
\put {$Q_2$} [b] at 150 10
\put {\small$SU(n_c)_1$} at 0 0
\put {\small$SU(n_c)_2$} at 100 0
\put {\scriptsize\stack{$SU(n_c)_{3}$,global}} at 200 0
\put {\scriptsize\stack{$SU(n_c)_{0}$,global}} at -100 0
\stpltsmbl
\tarrow from -75 0 to -48 0
\plot -48 0 -25 0 /
\tarrow from 25 0 to 52 0
\plot 52 0 75 0 /
\tarrow from 125 0 to 152 0
\plot 152 0 175 0 /
\linethickness=0pt
\putrule from -125 0 to 225 0
\endpicture$$
\caption{2-site chain}
\label{fig1e}
\end{center}
\end{figure}

However, before we descend to these technicalities, we will state one of our
principal results for the general invariants and use it to do a nontrivial
calculation. The basic splitting relation satisfied by these moduli spaces
can be written in terms of determinants of strings of contiguous $Q$'s as (a
determinant with no argument is defined to be $1$)
\begin{equation}
\begin{array}{c}
\det(Q_i\cdots Q_k)\,
\det(Q_j\cdots Q_\ell)
-\det(Q_i\cdots Q_\ell)\,
\det(Q_j\cdots Q_k)\\
\displaystyle
=\det(Q_i\cdots Q_{j-2})\,
\det(Q_{k+2}\cdots Q_\ell)\,
\prod_{m=j}^{k+1}\Lambda_m^{2n_c}\\
\mbox{for  $i\leq j-1$, $j\leq k+1$, and
$k\leq\ell-1$}.
\end{array}
\label{splitting}
\end{equation}
To attempt to convince the reader that there is something interesting going
on in this relation, we will apply it in an example that is simple but
nontrivial, the 2-site chain shown in fig.~\ref{fig1e}.

Applying the splitting relation, (\ref{splitting}), for $i=0$, $j=k=1$ and
$\ell=2$ gives
\begin{equation}
\det(Q_0Q_1)\,
\det(Q_1Q_2)
-\det(Q_0Q_1Q_2)\,
\det(Q_1)
=
\Lambda_1^{2n_c}\,
\Lambda_2^{2n_c}.
\label{ex-splitting}
\end{equation}
We will go very far out in moduli space in
the classical $D$-flat direction
\begin{equation}
Q_0=Q_2=
\pmatrix{
A_{m_c}&0\cr
0&0\cr
}
\quad\mbox{and}\quad
Q_1=\pmatrix{
\sqrt{v^2I_{m_c}+|A_{m_c}|^2}&0\cr
0&vI_{\ell_c}\cr
}
\label{ex-qjs}
\end{equation}
where $I_{m_c}$ and $I_{\ell_c}$ are $m_c\times m_c$ and
$\ell_c\times\ell_c$ identity
matrices with $m_c+\ell_c=n_c$, and
$A_{m_c}$ is a diagonal $m_c\times m_c$ matrix. In perturbation theory,
this breaks the symmetry down to a single $SU(\ell_c)$ and produces a QCD-like
effective low energy theory with a QMMS. With the splitting relations, we
will be able to write down the $\Lambda$ parameter in the low energy theory
in a single step. In this case, the calculation is simple enough that one
can check it perturbatively, but it is quite non-trivial.

In perturbation theory, the
fields in the low energy theory are effectively frozen at their vacuum
values except for the ``quark fields'' of the unbroken $SU(\ell_c)$ and we
can go over to the low energy theory by
replacing the high energy fields by
\begin{equation}
Q_0=
\pmatrix{
A_{m_c}&0\cr
0&\tilde Q_0\cr
}
\quad
Q_1=\pmatrix{
\sqrt{v^2I_{m_c}+|A_{m_c}|^2}&0\cr
0&vI_{\ell_c}\cr
}
\quad
Q_2=
\pmatrix{
A_{m_c}&0\cr
0&\tilde Q_2\cr
}.
\label{ex-qjs2}
\end{equation}
Inserting (\ref{ex-qjs2}) into (\ref{ex-splitting}), we find the
splitting condition for the low energy theory,
\begin{equation}
\det(\tilde Q_0\tilde Q_{2})
=
\det(\tilde Q_0)\det(\tilde Q_{2})-{\Lambda_1^{2n_c}\,\Lambda_2^{2n_c}\over
\det(A_{m_c}^2)\det(v^2I_{m_c}+|A_{m_c}|^2)\,v^{2\ell_c}}.
\label{ex-plr1}
\end{equation}
But this is the QMMS condition,
(\ref{qmmsnc}), for the 1-site $SU(\ell_c)$ theory.
Evidently, the $\tilde\Lambda$ parameter in the low energy theory is given
by
\begin{equation}
\tilde\Lambda^{2\ell_c}={\Lambda_1^{2n_c}\,\Lambda_2^{2n_c}\over
\det(A_{m_c}^2)\det(v^2I_{m_c}+|A_{m_c}|^2)\,v^{2\ell_c}}.
\label{ex-plr2}
\end{equation}
While it is possible to obtain this result directly using perturbation
theory\footnote{One has to keep in mind that the $\Lambda$ parameters depend upon the 
{\bf holomorphic} coupling.  In order to compare with perturbation theory, one has to 
relate the holomorphic coupling to the canonical coupling.  See \cite{Arkani-Hamed:1997mj}
for a nice discussion on this matter.}, 
the splitting relations are a {\bf much} simpler way of doing the
analysis. We hope that this will encourage the reader to wade through the
detailed analysis of $SU(2)$ models in sections
\ref{su2a}-\ref{checks}. However, those impatient for more applications to
moose chain models can
skip directly to section \ref{ssb} on page \pageref{ssb}.

\section{$SU(2)$ Analysis\label{su2a}}
Before we begin, we should explain why we are first analyzing the case
$SU(2)^N$
instead of the more general case of $SU(n_c)^N$.  We do this not only for
simplicity, but because it is actually possible to add gauge invariant mass
terms in the superpotential 
for the $Q$'s which are not allowed for larger $n_c$.  This is just due
to the fact that $\det (Q)$ is a mass term only for $n_c = 2$.  On the
other hand for $n_c > 2$, no such gauge invariant mass terms exist.
In fact, though we have much more control in the $SU(2)$ case, we expect
the results we obtain for the $SU(2)$ chains to be valid
for $SU(n_c)$ chains as well, and we will discuss the evidence for this later.

Overall, our methodology, assumptions, and general notation follow
that of Intriligator's ``integrating in'' method \cite{Intriligator:1994uk}.  The method is
to introduce mass terms that reduce our theory to one whose low energy
superpotential is known.
Then to integrate back in the field and deduce the original superpotential,
one just performs the
inverse Legendre transformation.  We will also abbreviate the notation for
determinants, defining
\begin{equation}
\d Q \equiv\det Q.
\end{equation}

\subsection{N=2 Chain}
Let's start with the smallest nontrivial
model, the N=2 chain.\footnote{Note:  This derivation mirrors that of the
original reference \cite{Intriligator:1994uk}, recast in our own language.}  If we add a mass term
for $Q_1$ the low energy theory will contain two disconnected $SU(2)$ gauge
theories with $n_f=1$.  The superpotentials for these theories are known
due to the Finnell-Pouliot calculation \cite{Finnell:1995dr}.  The terms we add
are
\begin{equation}
W_{tree} = m \: \d Q_1 + \tr(\lambda Q_0Q_1Q_2)
\label{eq:tree}
\end{equation}
where $\lambda$ is a 2 by 2 matrix.\footnote{The `integrating in' method requires
couplings in $W_{tree}$ to all gauge invariants involving the massive field $Q_1$, which is
why we've added the trace term.}

To integrate out $Q_1$, we solve the equation of motion
\begin{equation}
0 = (dW_{tree}/dQ_1)^T = m(Q_1)^{-1}\,\d Q_1 + Q_2\lambda Q_0
\end{equation}
and get
\begin{equation}
W_{tree,d} = -m\: \d Q_1 = -\frac{\d(Q_2\lambda Q_0)}{m} = -\frac{\d\lambda
\: \d Q_0 \: \d Q_2}{m}.
\label{eq:treed}
\end{equation}
Note that in the last equality, we have assumed that the large determinant
breaks up into the three separate determinants.  This is justified since
$Q_0$ and $Q_2$ are not charged under a common gauge group, and thus we are
only breaking up indices with no gauge dynamics between them.  This point
is especially important when this is generalized to larger N.

Now, we know the low energy superpotential of the 2 disconnected 1-site
theories is
\begin{equation}
W_d = \frac{\Lambda_{1,d}^5}{\d Q_0} + \frac{\Lambda_{2,d}^5}{\d Q_2} =
      \frac{m\Lambda_1^4}{\d Q_0} + \frac{m\Lambda_2^4}{\d Q_2}
\end{equation}
where we have assumed the matching conditions
\begin{equation}
\Lambda_{i,d}^5 = m\Lambda_i^4 \; \; {\rm for} \; \; i = 1, 2.
\label{eq:matching}
\end{equation}
The matching conditions are true if the gauge couplings of the high and low
energy theories match at the scale $m$ and the $\theta$ parameters are
equal.  The matching occurs without threshold corrections in
$\overline{DR}$.

So to integrate $Q_1$ back in, we perform the inverse Legendre
transform. We do this by integrating $m$ and $\lambda$ out of $$W_n = W_d +
W_{tree,d} - W_{tree}$$
\begin{equation}
=  \frac{m\Lambda_1^4}{\d Q_0} + \frac{m\Lambda_2^4}{\d Q_2}
-\frac{\d\lambda \,\d Q_0 \,\d Q_2}{m} - m\,\d Q_1 - \tr(\lambda Q_0Q_1Q_2).
\end{equation}
$$ $$
$\lambda$'s equation of motion is
\begin{equation}
0 = (dW_n/d\lambda)^T = -\frac{\d Q_0 \,\d Q_2}{m}\d\lambda \; \lambda^{-1}
- Q_0Q_1Q_2
\end{equation}
and substituting this back into $W_n$ gives
\begin{equation}
W_u = m\left ( \frac{\Lambda_1^4}{\d Q_0} + \frac{\Lambda_2^4}{\d Q_2} - \d Q_1
        + \frac{\d(Q_0Q_1Q_2)}{\d Q_0 \,\d Q_2}  \right ).
\end{equation}
Integrating out $m$ gives a zero superpotential for the high energy theory
and also enforces the constraint
\begin{equation}
\d(Q_0Q_1Q_2) = \d Q_0\,\d Q_1\,\d Q_2 - \Lambda_1^4 \,\d Q_2 - \Lambda_2^4
\,\d Q_0.
\end{equation}
Thus, there is a quantum modified moduli space and we have derived the
highest constraint for it when N = 2.  Notice that it has the form of the
classical constraint plus nonperturbative terms.  Now, as discussed before,
we only expected to get information on one constraint.  To deduce the two
remaining constraints for $\d(Q_0Q_1)$ and $\d(Q_1Q_2)$ we will use
indirect arguments that will be discussed later in the paper.

\begin{figure}[htb]
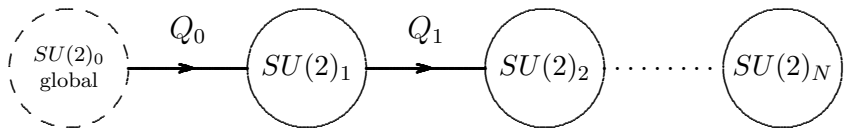

\begin{center}
$$\beginpicture
\setcoordinatesystem units <.9\tdim,.9\tdim>
\circulararc 360 degrees from 25 0 center at 0 0
\circulararc 360 degrees from 100 25 center at 100 0
\circulararc 360 degrees from 200 25 center at 200 0
\setdashes
\circulararc 360 degrees from -100 25 center at -100 0
\setsolid
\put {$Q_0$} [b] at -50 10
\put {$Q_1$} [b] at 50 10
\put {\small$SU(2)_1$} at 0 0
\put {\small$SU(2)_2$} at 100 0
\put {\small$SU(2)_N$} at 200 0
\put {\scriptsize\stack{$SU(2)_{0}$,global}} at -100 0
\stpltsmbl
\tarrow from -75 0 to -48 0
\plot -48 0 -25 0 /
\tarrow from 25 0 to 52 0
\plot 52 0 75 0 /
\setdots
\plot 125 0 175 0 /
\linethickness=0pt
\putrule from -125 0 to 225 0
\endpicture$$
\caption{N-site chain with a missing end link}
\label{fig2}
\end{center}
\end{figure}

Now, we can start an iterative process to determine the highest constraint
for the larger N models.  Given the highest constraint for an N-site chain,
give a mass term to $Q_N$ and integrate out $Q_N$ to discover the
superpotential for an N-site chain with one missing end link (see
fig.~\ref{fig2}).  Once we have this superpotential, we can take a
(N+1)-site chain and give a mass term to it's $Q_N$ (along with the
analogous trace term we put in above for N=2), and then the lower theory
consists of the N-site chain with a missing end link plus a 1-site chain
with a missing link.  Both of those superpotentials are now known, so we
can just integrate back in $Q_N$ to discover the highest constraint for the
(N+1)-site chain.  Inductively, this process can be repeated for all N.

\subsection{2-site Chain With A Missing End Link}
Let's see how we can integrate out an end link to determine the
superpotential of a 2-site chain with a missing end link.  From the added
mass term $W = m\: \d Q_2$, we can integrate out $Q_2$ by using the
constraint for $\d(Q_0Q_1Q_2)$ in the following way.  In the low energy
theory, we have the fields $Q_0$ and $Q_1$.  To integrate out $Q_2$ we
should think carefully on what set of independent fields we choose for the
high energy theory.  The obvious choice is the set $$\d Q_0, \; \d Q_1, \;
Q_0Q_1Q_2$$ since it contains a complete set for the low energy theory as
well.

To proceed, we must rewrite the mass term in terms of this independent set,
and that's where we use the constraint to get
\begin{equation}
m\: \d Q_2 = m \frac{\d(Q_0Q_1Q_2) + \Lambda_2^4 \d Q_0}{\d Q_0\d Q_1 -
\Lambda_1^4}.
\end{equation}
Now, we must integrate out $Q_0Q_1Q_2$ since this is the only independent
in our set that
does not appear in the low energy theory.  This
just sets $Q_0Q_1Q_2 = 0$, which leaves
\begin{equation}
W = \frac{m\Lambda_2^4 \,\d Q_0}{\d Q_0\,\d Q_1 - \Lambda_1^4} =
 \frac{\Lambda_{2,d}^5 \,\d Q_0}{\d Q_0\,\d Q_1 - \Lambda_1^4}.
\label{eq:2end}
\end{equation}
This is precisely the result obtained by Seiberg, {\it et al.\/} \cite{Intriligator:1994jr},
where we used the matching condition (\ref{eq:matching}) in the last step
of (\ref{eq:2end}).
This method can be applied for any N.\footnote{Alternatively, if we start with the final QMMS
constraints, by using a superpotential with the mass term $m\; \d Q_2$ and Lagrange multipliers
enforcing the constraints, we can derive the same superpotential by integrating out $\d Q_2,
\d(Q_1 Q_2), {\rm and}\; Q_0Q_1Q_2$.  Thus, this procedure is internally consistent.}

\section{Results\label{results}}
Now that we can determine the superpotentials for a chain with a missing
end link, it is straightforward to determine the highest
QMMS constraints for all N-site chains. In section~\ref{ssb} and in the
appendix, we will discuss evidence that this result for the highest
constraint generalizes to the lower constraints in an entirely trivial way.
The lower constraint on any contiguous set of link fields is obtained by
simply ignoring the other links and sites. In other words, the constraints
are local in theory space. The constraints have the form
$$\d(Q_i \cdots Q_j) =$$
\begin{equation}
\d Q_i\,\d Q_{i+1}\cdots \d Q_j
+ \sum_{\rm \stackrel{\scriptstyle any\;number}{\stackrel{\scriptstyle of\;
neighbor}{contractions}}} \,\d Q_i\cdots
 \overbrace{\d Q_{k-1}\,\d Q_k}^{-\Lam{k}}\cdots \d Q_j \equiv
\D(Q_i,\cdots ,Q_j)
\label{eq:expectedform}
\end{equation}
where the rule is that once a set of neighbors is contracted the
corresponding $Q$'s are gone, so these
$Q$'s cannot be
contracted with their other neighbors.
So for instance, we would have
\begin{equation}
\begin{array}{c}
\d(Q_0Q_1Q_2Q_3) = \D(Q_0,Q_1,Q_2,Q_3)\\
=\d Q_0\,\d Q_1\,\d Q_2\,\d Q_3 -\Lam{1}\,\d Q_2\,\d Q_3 -\Lam{2}\,\d
Q_0\,\d Q_3
-\Lam{3}\,\d Q_0\,\d Q_1 + \Lam{1}\Lam{3}\,.
\end{array}
\end{equation}

The $\D$ functions that appear in these constraints (defined in
(\ref{eq:expectedform})) satisfy the remarkable splitting property already
discussed in section~\ref{teaser}:
Let us abbreviate
\begin{equation}
\cD_{i,j}\equiv \D(Q_i,\cdots,Q_j)\,.
\label{curlyd}
\end{equation}
For convenience, we will also define
\begin{equation}
\cD_{i,i-1}=\mbox{``$\D()$''}\equiv1
\quad\mbox{and}\quad\cD_{i,j}=0\;\mbox{for}\;
j<i-1\,.
\label{convenience}
\end{equation}
The fundamental recursion relations (trivially derivable from the
definitions, (\ref{eq:expectedform})) are
\begin{equation}
\cD_{i,j}=\cD_{i,j-1}\,\d Q_{j}-\cD_{i,j-2}\,\Lambda^4_{j}
\quad\mbox{and}\quad
\cD_{i,j}=\d Q_{i}\,\cD_{i+1,j}-\Lambda^4_{i+1}\,\cD_{i+2,j}\,.
\label{rec1}
\end{equation}

With the definition (\ref{convenience}), (\ref{rec1}) is valid for all
$i\leq j$.
From these one can derive
the general
splitting relation, for $i\leq j-1$, $j\leq k+1$, and $k\leq\ell-1$
\begin{equation}
\cD_{i,k}\,\cD_{j,\ell}
-\cD_{i,\ell}\,\cD_{j,k}
=\cD_{i,j-2}\,\cD_{k+2,\ell}\,\prod_{m=j}^{k+1}\Lambda_m^4
\label{eq:splitting}
\end{equation}
We call these splitting relations because the product of $\Lambda$'s that
appears on the right hand side is associated in a rather direct way with
the splitting of the determinants of products of link fields.
The point is that we can think of (\ref{eq:splitting}) in two different
ways. Using (\ref{curlyd}) and the definition of the $\D$ functions,
(\ref{eq:splitting}) is simply an identity. But using the constraints,
\begin{equation}
\cD_{i,j}= \d(Q_i,\cdots,Q_j)\,,
\label{dproducts}
\end{equation}
(\ref{eq:splitting}) becomes a dynamical statement about the determinants
of the products.
If any of the $\Lambda$'s in
the product goes to zero, we can split each of the determinants into two
pieces in such a way that the two terms on the left hand side of
(\ref{eq:splitting}) cancel. For example, suppose $\Lambda_m\to0$ for
$j< m< k+1$. This $\Lambda$ characterizes the interaction between the links
$Q_m$ and $Q_{m+1}$. If $\Lambda_m\to0$, then the index shared by these two
fields becomes a global index, and the determinant of a product of fields
can be {\bf split} at that index into the product of determinants:
\begin{equation}
\begin{array}{c}
\det(Q_i\cdots Q_k)
\to
\det(Q_i\cdots Q_m)
\det(Q_{m+1}\cdots Q_k)
\\
\det(Q_j\cdots Q_\ell)
\to
\det(Q_j\cdots Q_m)
\det(Q_{m+1}\cdots Q_\ell)
\\
\det(Q_i\cdots Q_\ell)
\to
\det(Q_i\cdots Q_m)
\det(Q_{m+1}\cdots Q_\ell)
\\
\det(Q_j\cdots Q_k)
\to
\det(Q_j\cdots Q_m)
\det(Q_{m+1}\cdots Q_k)
\end{array}
\label{explicitsplitting}
\end{equation}
If we put (\ref{explicitsplitting}) into the left hand side of
(\ref{eq:splitting}), the two terms cancel.

We will also see that the splitting relation (\ref{eq:splitting}) incorporates much of the
perturbative physics we expect of the chain.
This is interesting because if one instead started with (\ref{convenience})
and (\ref{eq:splitting}),
one could derive the forms of the D's (\ref{eq:expectedform}).  Therefore,
the given constraints are the unique set that gives the physics contained within
the splitting relation, which gives us confidence that we have found the correct constraints.

\begin{figure}[htb]
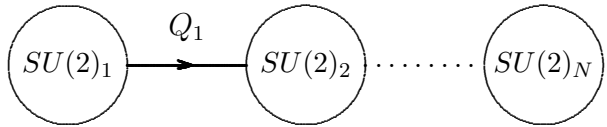

\begin{center}
$$\beginpicture
\setcoordinatesystem units <.9\tdim,.9\tdim>
\circulararc 360 degrees from 25 0 center at 0 0
\circulararc 360 degrees from 100 25 center at 100 0
\circulararc 360 degrees from 200 25 center at 200 0
\setsolid
\put {$Q_1$} [b] at 50 10
\put {\small$SU(2)_1$} at 0 0
\put {\small$SU(2)_2$} at 100 0
\put {\small$SU(2)_N$} at 200 0
\stpltsmbl
\tarrow from 25 0 to 52 0
\plot 52 0 75 0 /
\setdots
\plot 125 0 175 0 /
\linethickness=0pt
\putrule from -25 0 to 225 0
\endpicture$$
\caption{N-site chain with no ends}
\label{fig3}
\end{center}
\end{figure}

In the process of finding the constraints, (\ref{eq:expectedform}), we've
discovered the following superpotentials:
for the N-site chain with a missing end as shown in fig.~\ref{fig2},
there is a generated superpotential
\begin{equation}
W = \frac{\Lambda_N^5 \,\D(Q_0,Q_1, \cdots ,Q_{N-2})}{\D(Q_0,Q_1, \cdots
,Q_{N-1})}
\label{eq:Nend}
\end{equation}
\noindent
and for the N-site chain with its two end links missing (see fig.~\ref{fig3}),
there is a superpotential
\begin{equation}
W = \frac{\Lambda_1^5 \,\D(Q_2,Q_3, \cdots, Q_{N-1}) +
          \Lambda_N^5 \,\D(Q_1,Q_2, \cdots, Q_{N-2})
          \pm 2\sqrt{\Lambda_1^5\Lambda_2^4\cdots \Lambda_{N-1}^4\Lambda_N^5}}
{\D(Q_1,Q_2, \cdots, Q_{N-1})}.
\label{eq:noend}
\end{equation}
$$ $$
This last superpotential also agrees with the known result (for N=2
\cite{Intriligator:1994jr}) with similar interpretations of the terms as contributions
both from instantons and gaugino condensation in the various $SU(2)$
groups.

\section{Consistency Checks\label{checks}}
An easy consistency check we can perform is letting some $\Lambda_k$ go to
zero.  This is equivalent to setting $g_k = 0$, making $SU(2)_k$ a global
symmetry.  In this limit, we expect a chain to break up into two disjoint
chains that don't communicate.  This behavior should be seen in the
constraints and superpotentials we have derived.  Looking at the N-site
chain in this limit, we expect determinants to factorize in the following
way
\begin{equation}
\d(Q_i \cdots Q_j) \to \d(Q_i \cdots Q_{k-1})\,\d(Q_k \cdots Q_j)
\end{equation}
since the $Q_i\cdots Q_{k-1}$ and $Q_k\cdots Q_j$ do not interact any longer.
The constraints should also factorize in this limit, and by using the
splitting property
for the D functions (\ref{eq:splitting}), we see that
\begin{equation}
\cD_{i,k-1}\,\cD_{k,j}
-\cD_{i,j}
=\cD_{i,k-2}\,\cD_{k+1,j}\,\Lambda_k^4
\label{rec-factor}
\end{equation}
which immediately implies
\begin{equation}
\D(Q_i, \cdots, Q_j) \stackrel{\Lambda_k \to 0}{\to} \D(Q_i, \cdots,
Q_{k-1})\,\D(Q_k, \cdots, Q_j)
\end{equation}
as well.  Thus the constraints for the N-site chain factor appropriately
into the constraints of a (k-1)-site chain and the constraints of a
(N-k)-site chain.

We can check the superpotentials as well.  For an N-site chain with one
missing end link, if
we take the limit $\Lambda_k \to 0$, we get
\begin{equation}
W \to \frac{\Lambda_N^5 \,\D(Q_k, \cdots ,Q_{N-2})}{\D(Q_k, \cdots ,Q_{N-1})}.
\end{equation}
This is precisely the superpotential for a (N-k)-site chain with a missing
end link plus
the superpotential for a (k-1)-site chain (i.e. zero).
For the N-site chain with both ends missing, the limit gives
\begin{equation}
W \to \frac{\Lambda_1^5 \,\D(Q_2, \cdots ,Q_{k-1})}{\D(Q_1, \cdots ,Q_{k-1})}
+ \frac{\Lambda_N^5 \,\D(Q_k, \cdots ,Q_{N-2})}{\D(Q_k, \cdots ,Q_{N-1})}
\end{equation}
and this is precisely the superpotentials for a (k-1)-site chain with a
missing end link
and a (N-k)-site chain with a missing end link.

\newcommand{\link}[2]{\put {\beginpicture
\setcoordinatesystem units <.5\tdim,.5\tdim>
\arrow <5\tdim> [.3,.6] from -25 0 to 5 0
\plot 2 0 25 0 /
\linethickness=0pt
\putrule from -25 0 to 25 0
\endpicture} at #1 #2 }

\begin{figure}[htb]
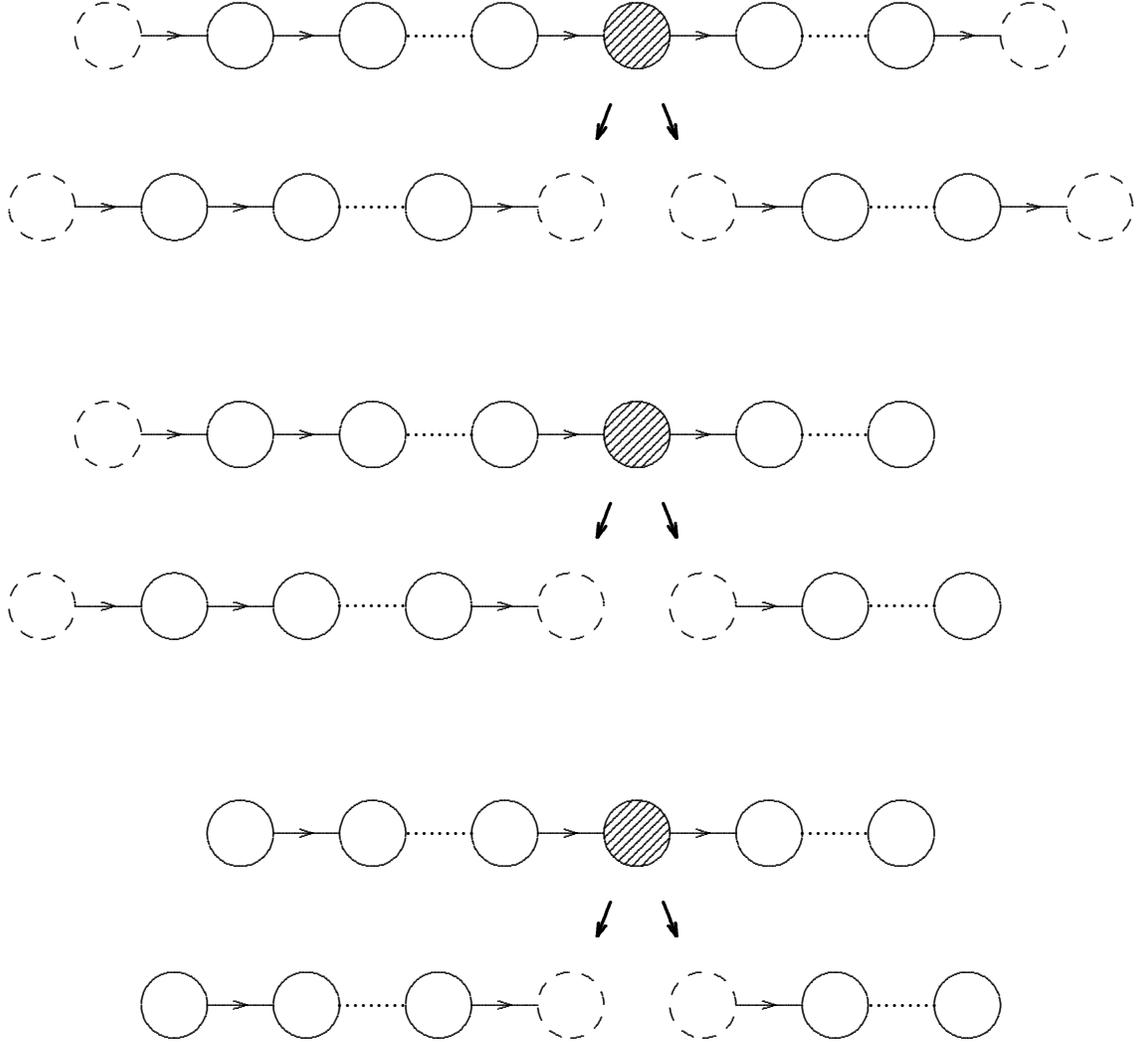

\begin{center}
$$\beginpicture
\setcoordinatesystem units <.5\tdim,.5\tdim>
\startrotation by .707 .707 about 0 0
\plot -25 0 25 0 /
\plot -24.5 5 24.5 5 /
\plot -24.5 -5 24.5 -5 /
\plot -22.9 10 22.9 10 /
\plot -22.9 -10 22.9 -10 /
\plot -20 15 20 15 /
\plot -20 -15 20 -15 /
\plot -15 20 15 20 /
\plot -15 -20 15 -20 /
\stoprotation
\circulararc 360 degrees from 0 25 center at 0 0
\circulararc 360 degrees from 100 25 center at 100 0
\circulararc 360 degrees from 200 25 center at 200 0
\circulararc 360 degrees from -100 25 center at -100 0
\circulararc 360 degrees from -200 25 center at -200 0
\circulararc 360 degrees from -300 25 center at -300 0
\link{50}{0}
\link{250}{0}
\link{-50}{0}
\link{-250}{0}
\link{-350}{0}
\setdashes
\circulararc 360 degrees from -400 25 center at -400 0
\circulararc 360 degrees from 300 25 center at 300 0
\setsolid
\circulararc 360 degrees from 150 -155 center at 150 -130
\circulararc 360 degrees from 250 -155 center at 250 -130
\circulararc 360 degrees from -150 -155 center at -150 -130
\circulararc 360 degrees from -250 -155 center at -250 -130
\circulararc 360 degrees from -350 -155 center at -350 -130
\link{100}{-130}
\link{300}{-130}
\link{-100}{-130}
\link{-300}{-130}
\link{-400}{-130}
\setdashes
\circulararc 360 degrees from 50 -155 center at 50 -130
\circulararc 360 degrees from -50 -155 center at -50 -130
\circulararc 360 degrees from -450 -155 center at -450 -130
\circulararc 360 degrees from 350 -155 center at 350 -130
\stpltsmbl
\setsolid
\tarrow from 20 -52 to 30 -78
\tarrow from -20 -52 to -30 -78
\setdots <3\tdim>
\plot 125 0 175 0 /
\plot -125 0 -175 0 /
\plot 225 -130 175 -130 /
\plot -225 -130 -175 -130 /
\linethickness=0pt
\putrule from -450 0 to 350 0
\putrule from 0 50 to 0 -200
\endpicture$$

$$\beginpicture
\setcoordinatesystem units <.5\tdim,.5\tdim>
\startrotation by .707 .707 about 0 0
\plot -25 0 25 0 /
\plot -24.5 5 24.5 5 /
\plot -24.5 -5 24.5 -5 /
\plot -22.9 10 22.9 10 /
\plot -22.9 -10 22.9 -10 /
\plot -20 15 20 15 /
\plot -20 -15 20 -15 /
\plot -15 20 15 20 /
\plot -15 -20 15 -20 /
\stoprotation
\circulararc 360 degrees from 0 25 center at 0 0
\circulararc 360 degrees from 100 25 center at 100 0
\circulararc 360 degrees from 200 25 center at 200 0
\circulararc 360 degrees from -100 25 center at -100 0
\circulararc 360 degrees from -200 25 center at -200 0
\circulararc 360 degrees from -300 25 center at -300 0
\link{50}{0}
\link{-50}{0}
\link{-250}{0}
\link{-350}{0}
\setdashes
\circulararc 360 degrees from -400 25 center at -400 0
\setsolid
\circulararc 360 degrees from 150 -155 center at 150 -130
\circulararc 360 degrees from 250 -155 center at 250 -130
\circulararc 360 degrees from -150 -155 center at -150 -130
\circulararc 360 degrees from -250 -155 center at -250 -130
\circulararc 360 degrees from -350 -155 center at -350 -130
\link{100}{-130}
\link{-100}{-130}
\link{-300}{-130}
\link{-400}{-130}
\setdashes
\circulararc 360 degrees from 50 -155 center at 50 -130
\circulararc 360 degrees from -50 -155 center at -50 -130
\circulararc 360 degrees from -450 -155 center at -450 -130
\stpltsmbl
\setsolid
\tarrow from 20 -52 to 30 -78
\tarrow from -20 -52 to -30 -78
\setdots <3\tdim>
\plot 125 0 175 0 /
\plot -125 0 -175 0 /
\plot 225 -130 175 -130 /
\plot -225 -130 -175 -130 /
\linethickness=0pt
\putrule from -450 0 to 350 0
\putrule from 0 50 to 0 -200
\endpicture$$

$$\beginpicture
\setcoordinatesystem units <.5\tdim,.5\tdim>
\startrotation by .707 .707 about 0 0
\plot -25 0 25 0 /
\plot -24.5 5 24.5 5 /
\plot -24.5 -5 24.5 -5 /
\plot -22.9 10 22.9 10 /
\plot -22.9 -10 22.9 -10 /
\plot -20 15 20 15 /
\plot -20 -15 20 -15 /
\plot -15 20 15 20 /
\plot -15 -20 15 -20 /
\stoprotation
\circulararc 360 degrees from 0 25 center at 0 0
\circulararc 360 degrees from 100 25 center at 100 0
\circulararc 360 degrees from 200 25 center at 200 0
\circulararc 360 degrees from -100 25 center at -100 0
\circulararc 360 degrees from -200 25 center at -200 0
\circulararc 360 degrees from -300 25 center at -300 0
\link{50}{0}
\link{-50}{0}
\link{-250}{0}
\setsolid
\circulararc 360 degrees from 150 -155 center at 150 -130
\circulararc 360 degrees from 250 -155 center at 250 -130
\circulararc 360 degrees from -150 -155 center at -150 -130
\circulararc 360 degrees from -250 -155 center at -250 -130
\circulararc 360 degrees from -350 -155 center at -350 -130
\link{100}{-130}
\link{-100}{-130}
\link{-300}{-130}
\setdashes
\circulararc 360 degrees from -50 -155 center at -50 -130
\circulararc 360 degrees from 50 -155 center at 50 -130
\stpltsmbl
\setsolid
\tarrow from 20 -52 to 30 -78
\tarrow from -20 -52 to -30 -78
\setdots <3\tdim>
\plot 125 0 175 0 /
\plot -125 0 -175 0 /
\plot 225 -130 175 -130 /
\plot -225 -130 -175 -130 /
\linethickness=0pt
\putrule from -450 0 to 350 0
\putrule from 0 50 to 0 -200
\endpicture$$
\caption{The consistency checks for $\Lambda_k\to0$. The filled
gauge group is the one for which $\Lambda\to0$.}
\label{fig4}
\end{center}
\end{figure}

\begin{figure}[htb]
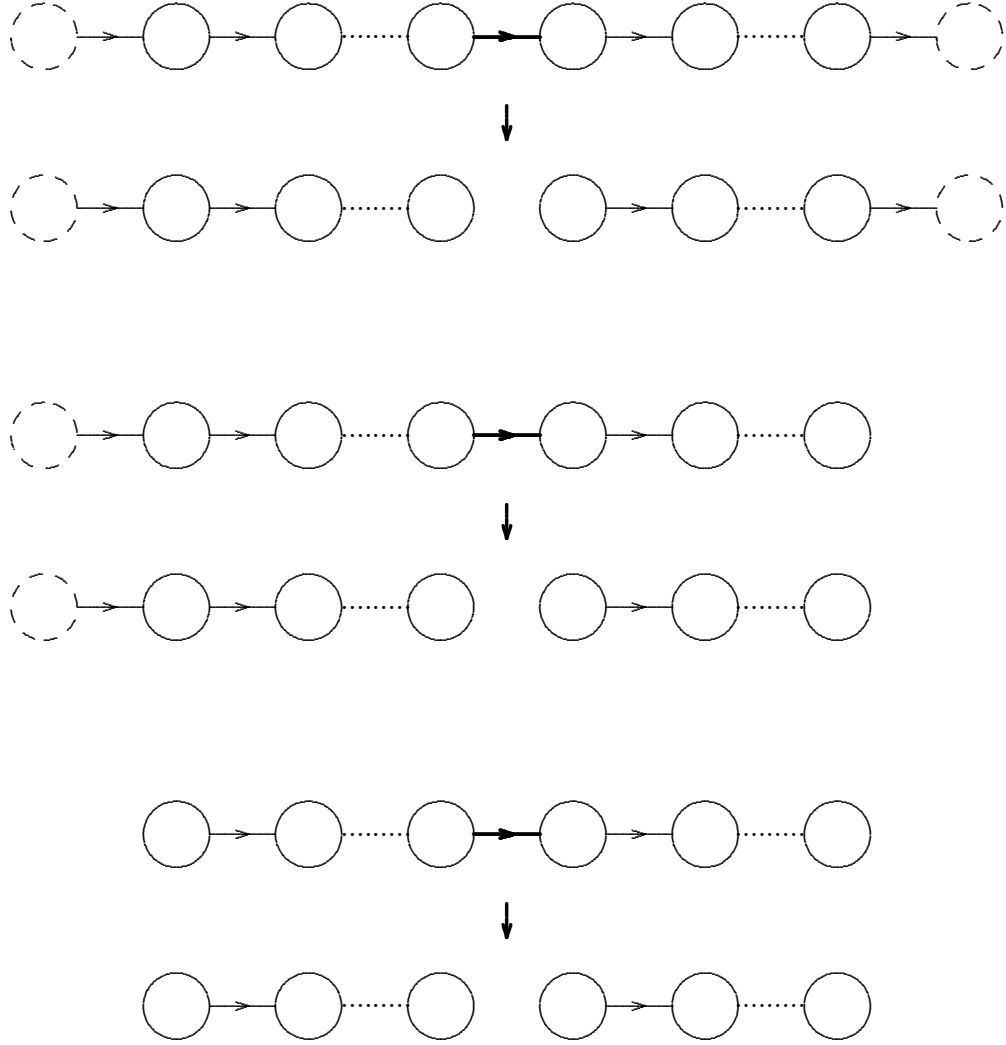

\begin{center}
$$\beginpicture
\setcoordinatesystem units <.5\tdim,.5\tdim>
\circulararc 360 degrees from 0 25 center at 0 0
\circulararc 360 degrees from 100 25 center at 100 0
\circulararc 360 degrees from 200 25 center at 200 0
\circulararc 360 degrees from -100 25 center at -100 0
\circulararc 360 degrees from -200 25 center at -200 0
\circulararc 360 degrees from -300 25 center at -300 0
\link{50}{0}
\link{250}{0}
\link{-250}{0}
\link{-350}{0}
\setdashes
\circulararc 360 degrees from -400 25 center at -400 0
\circulararc 360 degrees from 300 25 center at 300 0
\setsolid
\circulararc 360 degrees from 0 -155 center at 0 -130
\circulararc 360 degrees from 100 -155 center at 100 -130
\circulararc 360 degrees from 200 -155 center at 200 -130
\circulararc 360 degrees from -100 -155 center at -100 -130
\circulararc 360 degrees from -200 -155 center at -200 -130
\circulararc 360 degrees from -300 -155 center at -300 -130
\link{50}{-130}
\link{250}{-130}
\link{-250}{-130}
\link{-350}{-130}
\setdashes
\circulararc 360 degrees from -400 -155 center at -400 -130
\circulararc 360 degrees from 300 -155 center at 300 -130
\setsolid
\stpltsmbl
\tarrow from -50 -52 to -50 -78
\setdots <3\tdim>
\plot 125 0 175 0 /
\plot -125 0 -175 0 /
\plot 125 -130 175 -130 /
\plot -125 -130 -175 -130 /
\setsolid
\setplotsymbol ({\large.})
\link{-50}{0}
\linethickness=0pt
\putrule from -450 0 to 350 0
\putrule from 0 50 to 0 -200
\endpicture$$

$$\beginpicture
\setcoordinatesystem units <.5\tdim,.5\tdim>
\circulararc 360 degrees from 0 25 center at 0 0
\circulararc 360 degrees from 100 25 center at 100 0
\circulararc 360 degrees from 200 25 center at 200 0
\circulararc 360 degrees from -100 25 center at -100 0
\circulararc 360 degrees from -200 25 center at -200 0
\circulararc 360 degrees from -300 25 center at -300 0
\link{50}{0}
\link{-250}{0}
\link{-350}{0}
\setdashes
\circulararc 360 degrees from -400 25 center at -400 0
\setsolid
\circulararc 360 degrees from 0 -155 center at 0 -130
\circulararc 360 degrees from 100 -155 center at 100 -130
\circulararc 360 degrees from 200 -155 center at 200 -130
\circulararc 360 degrees from -100 -155 center at -100 -130
\circulararc 360 degrees from -200 -155 center at -200 -130
\circulararc 360 degrees from -300 -155 center at -300 -130
\link{50}{-130}
\link{-250}{-130}
\link{-350}{-130}
\setdashes
\circulararc 360 degrees from -400 -155 center at -400 -130
\setsolid
\stpltsmbl
\tarrow from -50 -52 to -50 -78
\setdots <3\tdim>
\plot 125 0 175 0 /
\plot -125 0 -175 0 /
\plot 125 -130 175 -130 /
\plot -125 -130 -175 -130 /
\setsolid
\setplotsymbol ({\large.})
\link{-50}{0}
\linethickness=0pt
\putrule from -450 0 to 350 0
\putrule from 0 50 to 0 -200
\endpicture$$

$$\beginpicture
\setcoordinatesystem units <.5\tdim,.5\tdim>
\circulararc 360 degrees from 0 25 center at 0 0
\circulararc 360 degrees from 100 25 center at 100 0
\circulararc 360 degrees from 200 25 center at 200 0
\circulararc 360 degrees from -100 25 center at -100 0
\circulararc 360 degrees from -200 25 center at -200 0
\circulararc 360 degrees from -300 25 center at -300 0
\link{50}{0}
\link{-250}{0}
\setsolid
\circulararc 360 degrees from 0 -155 center at 0 -130
\circulararc 360 degrees from 100 -155 center at 100 -130
\circulararc 360 degrees from 200 -155 center at 200 -130
\circulararc 360 degrees from -100 -155 center at -100 -130
\circulararc 360 degrees from -200 -155 center at -200 -130
\circulararc 360 degrees from -300 -155 center at -300 -130
\link{50}{-130}
\link{-250}{-130}
\setsolid
\stpltsmbl
\tarrow from -50 -52 to -50 -78
\setdots <3\tdim>
\plot 125 0 175 0 /
\plot -125 0 -175 0 /
\plot 125 -130 175 -130 /
\plot -125 -130 -175 -130 /
\setsolid
\setplotsymbol ({\large.})
\link{-50}{0}
\linethickness=0pt
\putrule from -450 0 to 350 0
\putrule from 0 50 to 0 -200
\endpicture$$

\caption{The consistency checks for integrating out a massive link. The
bold link is the
massive link.}
\label{fig5}
\end{center}
\end{figure}

Another consistency check is that adding a mass term for a $Q_k$ and
integrating out that $Q_k$ does properly flow down, upon using the matching
condition (\ref{eq:matching}).  We've already shown how this works for the
N-site chain, so let's see how the superpotentials reduce when we integrate
out the quark superfield.\footnote{In a similar argument about choosing the
correct set of independent gauge invariants, to integrate out $Q_k$, we
just integrate out $\d Q_k$.}  For the N-site chain with a missing end
link, integrating out the quark gives the superpotential
$$
W_{\rm low \ energy} =  \frac{m\Lambda_k^4 \,\D(Q_0, \cdots
,Q_{k-2})}{\D(Q_0, \cdots ,Q_{k-1})}$$
\begin{equation}
+ \frac{m\Lambda_{k+1}^4 \,\D(Q_{k+2}, \cdots, Q_{N-1}) +
          \Lambda_N^5 \,\D(Q_{k+1}, \cdots, Q_{N-2})
          \pm 2\sqrt{m\Lambda_{k+1}^4\cdots \Lambda_{N-1}^4\Lambda_N^5}}
{\D(Q_{k+1}, \cdots, Q_{N-1})}
\end{equation}
which is just the superpotentials for a k-site chain with a missing end
link and a
(N-k)-site chain missing both end links.
Finally, for the N-site chain missing both end links, integrating out $Q_k$
gives
$$
W_{\rm low \ energy} =  \frac{\Lambda_1^5 \,\D(Q_2, \cdots, Q_{k-1}) +
          m\Lambda_k^4 \,\D(Q_1, \cdots, Q_{k-2})
          \pm 2\sqrt{m\Lambda_1^5\Lambda_2^4\cdots \Lambda_k^4}}
{\D(Q_1, \cdots, Q_{k-1})}
$$
\begin{equation}
+ \frac{m\Lambda_{k+1}^4 \,\D(Q_{k+2}, \cdots, Q_{N-1}) +
          \Lambda_N^5 \,\D(Q_{k+1}, \cdots, Q_{N-2})
          \pm 2\sqrt{m\Lambda_{k+1}^4\cdots \Lambda_{N-1}^4\Lambda_N^5}}
{\D(Q_{k+1}, \cdots, Q_{N-1})}
\end{equation}
and this is just the superpotentials for a k-site chain with both end links
missing and a
(N-k)-site chain with both ends links missing.
Thus, the results come out as expected, thanks to some nontrivial algebra,
the matching
conditions (\ref{eq:matching}), and use of the splitting relation
(\ref{eq:splitting}). For clarity, a graphical depiction of these two types of checks is
given in the figures~\ref{fig4}~and~\ref{fig5}.

\section{Spontaneous gauge symmetry breaking\label{ssb}}

In this section, we will see that by appropriately choosing parameters in
a Moose-chain model and by going to the relevant region of moduli space, we
can unravel the effect of spontaneous symmetry breaking in these models,
and derive exact results for the matching of gauge couplings in spontaneous
symmetry breaking.

For those who skimmed lightly over the previous three sections, let us
recapitulate the results for the QMMS conditions for the moose chain.
The basic constraint looks like
\begin{equation}
\begin{array}{c}
\d(Q_i \cdots Q_j) =
             \d Q_i\,\d Q_{i+1}\cdots \d Q_j
\\ \displaystyle
+ \sum_{\rm \stackrel{\scriptstyle any\;number}{\stackrel{\scriptstyle of\;
neighbor}{contractions}}} \,\d Q_i\cdots
 \overbrace{\d Q_{k-1}\,\d Q_k}^{-\Lam{k}}\cdots \d Q_j \equiv
\D(Q_i,\cdots ,Q_j)
\end{array}
\label{d-function}
\end{equation}
We often abbreviate
\begin{equation}
\cD_{i,j}\equiv \D(Q_i,\cdots,Q_j)\,,\quad
\cD_{i,i-1}=\mbox{``$\D()$''}\equiv1\,,
\quad\cD_{i,j}=0\;\mbox{for}\;
j<i-1\,.
\label{d-abbreviate}
\end{equation}
Then the general
splitting relation, for $i\leq j-1$, $j\leq k+1$, and $k\leq\ell-1$ is
\begin{equation}
\cD_{i,k}\,\cD_{j,\ell}
-\cD_{i,\ell}\,\cD_{j,k}
=\cD_{i,j-2}\,\cD_{k+2,\ell}\,\prod_{m=j}^{k+1}\Lambda_m^4.
\label{d-splitting}
\end{equation}

We now begin by considering spontaneous symmetry breaking in
an $SU(2)$ chain.
The basic idea is that when the $SU(2)_j\times\cdots\times SU(2)_{k+1}$
subgroup is
spontaneously broken by vacuum values of the fields,
$Q_j,\cdots, Q_k$ down to the diagonal $SU(2)_{j,k+1}$, the result should be
a theory described by another moose of the same general form, but with the
$j, \cdots, k+1$ sites collapsed into a single site. There should be some way of
choosing the fields in the collapsed theory to satisfy all the conditions
that we have found for the moduli space in theories of this general
form. We will see that this is in fact possible, and that it provides
interesting information on the matching relations between the two
theories.

\begin{figure}[htb]
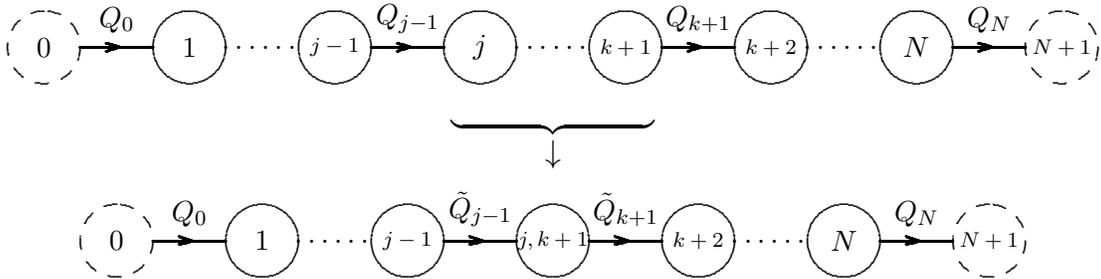

\begin{center}
$$\beginpicture
\setcoordinatesystem units <.55\tdim,.55\tdim>
\circulararc 360 degrees from 25 0 center at 0 0
\circulararc 360 degrees from 100 25 center at 100 0
\circulararc 360 degrees from 200 25 center at 200 0
\circulararc 360 degrees from 300 25 center at 300 0
\circulararc 360 degrees from 400 25 center at 400 0
\circulararc 360 degrees from 500 25 center at 500 0
\setdashes
\circulararc 360 degrees from -100 25 center at -100 0
\circulararc 360 degrees from 600 25 center at 600 0
\setsolid
\put {\small$Q_0$} [b] at -50 10
\put {\small$Q_{j-1}$} [b] at 150 10
\put {\small$Q_{k+1}$} [b] at 350 10
\put {\small$Q_N$} [b] at 550 10
\put {\small$0$} at -100 0
\put {\small$1$} at 0 0
\put {\scriptsize$j-1$} at 100 0
\put {\small$j$} at 200 0
\put {\scriptsize$k+1$} at 300 0
\put {\scriptsize$k+2$} at 400 0
\put {\small$N$} at 500 0
\put {\scriptsize$N+1$} at 600 0
\stpltsmbl
\tarrow from -75 0 to -48 0
\plot -48 0 -25 0 /
\tarrow from 125 0 to 152 0
\plot 152 0 175 0 /
\tarrow from 325 0 to 352 0
\plot 352 0 375 0 /
\tarrow from 525 0 to 552 0
\plot 552 0 575 0 /
\setdots
\plot 25 0 75 0 /
\plot 225 0 275 0 /
\plot 425 0 475 0 /
\linethickness=0pt
\putrule from -125 0 to 625 0
\endpicture$$
$$\underbrace{\beginpicture
\setcoordinatesystem units <.55\tdim,.55\tdim>
\linethickness=0pt
\putrule from -70 0 to 70 0
\putrule from 0 5 to 0 0
\endpicture}_{\displaystyle\downarrow}$$
$$\beginpicture
\setcoordinatesystem units <.55\tdim,.55\tdim>
\circulararc 360 degrees from 25 0 center at 0 0
\circulararc 360 degrees from 100 25 center at 100 0
\circulararc 360 degrees from 200 25 center at 200 0
\circulararc 360 degrees from 300 25 center at 300 0
\circulararc 360 degrees from 400 25 center at 400 0
\setdashes
\circulararc 360 degrees from -100 25 center at -100 0
\circulararc 360 degrees from 500 25 center at 500 0
\setsolid
\put {\small$Q_0$} [b] at -50 10
\put {\small$\tilde Q_{j-1}$} [b] at 150 10
\put {\small$\tilde Q_{k+1}$} [b] at 250 10
\put {\small$Q_N$} [b] at 450 10
\put {\small$0$} at -100 0
\put {\small$1$} at 0 0
\put {\scriptsize$j-1$} at 100 0
\put {\scriptsize$j,k+1$} at 200 0
\put {\scriptsize$k+2$} at 300 0
\put {\small$N$} at 400 0
\put {\scriptsize$N+1$} at 500 0
\stpltsmbl
\tarrow from -75 0 to -48 0
\plot -48 0 -25 0 /
\tarrow from 125 0 to 152 0
\plot 152 0 175 0 /
\tarrow from 225 0 to 252 0
\plot 252 0 275 0 /
\tarrow from 425 0 to 452 0
\plot 452 0 475 0 /
\setdots
\plot 25 0 75 0 /
\plot 325 0 375 0 /
\linethickness=0pt
\putrule from -125 0 to 525 0
\putrule from 0 0 to 0 -40
\endpicture$$
\caption{The low energy theory below a symmetry breaking scale}
\label{fig6}
\end{center}
\end{figure}

More precisely, we consider a set of parameters such that
$\Lambda_j,\cdots,\Lambda_{k+1}$ are large compared to the scale of our effective
theory, but we go far out in moduli space to make
\begin{equation}
\cD_{jk}=\D(Q_j,\cdots, Q_k)=\d(Q_j\cdots Q_k)
\end{equation}
very large, so that
\begin{equation}
{\Lambda^4_j\cdots\Lambda^4_{k+1}\over\D(Q_j,\cdots ,Q_k)^2}
\end{equation}
is small compared to $\Lambda^4_j,\cdots,\Lambda^4_{k+1}$.
The effective low energy theory should then describe a theory with the
$SU(2)_j\times\cdots\times SU(2)_{k+1}$ gauge subgroup replaced by the
diagonal, unbroken $SU(2)_{j,k+1}$ as shown in fig.~\ref{fig6}.

There is no problem analyzing this in perturbation theory if the non-zero
$\d Q_j$'s are large and the $\Lambda$'s are small. In this limit, we can
explicitly write down the vacuum value of the fields and work out the
details of the super-Higgs mechanism. But we want more than
that. We want to understand this in the gauge invariant language of the
QMMS conditions on the high energy and low energy theory. This will give us
detailed nonperturbative information.

Intuitively, what do we expect when the gauge groups break down to the
diagonal?  First of all, fields that do not transform under the breaking
groups should be unchanged upon going to the low energy theory.  This seems
reasonable due to the locality in theory space and the fact that these
fields do not experience the dynamics of the broken gauge groups.  On the
other hand, we don't expect fields that transform only under the 
broken groups (i.e. $Q_j, \cdots, Q_k$) to appear in the low energy fields
except within composites.  This is due to the complementarity 
of the Higgs and the confining phases, which holds since
the squarks are in the fundamental representations.

In fig.~\ref{fig6}, we have displayed these assumptions by leaving
$Q_0,\cdots,Q_{j-2}$ and $Q_{k+2},\cdots,Q_N$ unchanged and
labeling the links that are charged under the $SU(2)_{j,k+1}$
group as $\tilde Q_{j-1}$ and $\tilde Q_{k+1}$.  These newly
introduced tilde fields should satisfy some obvious constraints.
In particular, they should have all the right properties under the
unbroken symmetries in the low energy theory, and not transform
under any symmetries that have disappeared in going to the low
energy theory.  However, explicitly writing $\tilde Q_{j-1}$ and $\tilde
Q_{k+1}$ in terms of the high energy $Q$'s is not expected to be
holomorphic and requires detailed knowledge about the 
K\"{a}hler potential.  On the other hand, if we think in terms of gauge invariants,
we will not encounter these subtleties.
In fact, we'll see that everything will naturally be holomorphic 
and cleanly defined, which will aid in interpreting results. 

Now that we've refocused our attention to gauge invariants, it is simple to
relate the low energy gauge invariants in terms of the high energy ones.
It shouldn't come as a surprise that the symmetries alone do not uniquely
determine the relationship.  However, we also have the restriction that the
low energy gauge invariants satisfy their QMMS constraints.  This along
with the assumption that the untilded fields are unchanged uniquely
determines the map between gauge invariants.  Let us emphasize that it was
not at all trivial that there was any consistent map between gauge
invariants.  That one exists and appears to be unique gives further
evidence for the validity of the proposed constraints.

The gauge invariant quantities in the low energy theory that involve
$\tilde Q_{j-1}$ and $\tilde Q_{k+1}$ must
be chosen to satisfy
\begin{equation}
\d(Q_i\cdots\tilde Q_{j-1})
=\biggl(\D(Q_j,\cdots,Q_{k})\biggr)^{-1}\,\d(Q_i\cdots Q_{k})
\label{dtildes1}
\end{equation}
\begin{equation}
\d(\tilde Q_{k+1}\cdots Q_{\ell})
=\biggl(\D(Q_j,\cdots,Q_{k})\biggr)^{-1}\,\d(Q_j\cdots Q_{\ell})
\label{dtildes2}
\end{equation}
\begin{equation}
\d(Q_i\cdots\tilde Q_{j-1}\tilde Q_{k+1}\cdots Q_\ell)
=\biggl(\D(Q_j,\cdots,Q_{k})\biggr)^{-1}\,\d(Q_i\cdots Q_{\ell})
\label{dtildes4}
\end{equation}
\begin{equation}
Q_0\cdots\tilde Q_{j-1}\tilde Q_{k+1}\cdots Q_N
=\biggl(\D(Q_j,\cdots,Q_{k})\biggr)^{-1/2}\,Q_0\cdots Q_N
\label{dtildes3}
\end{equation}
Evidently, this incorporates all the right symmetry properties and is the unique
choice that satisfies the properties stated above. 
Notice that the factors of $\D(Q_j,\cdots,Q_{k})$ in
(\ref{dtildes1})-(\ref{dtildes3}) are associated entirely with the fields
$\tilde Q_{j-1}$ and $\tilde Q_{k+1}$.

We can rewrite (\ref{eq:splitting}) as
\begin{equation}
{\cD_{i,\ell}\over\cD_{j,k}}
={\cD_{i,k}\,\cD_{j,\ell}\over\cD_{j,k}^2}
-\cD_{i,j-2}\,\cD_{k+2,\ell}\,{\prod_{m=j}^{k+1}\Lambda_m^4\over \cD_{j,k}^2}.
\label{eq:splitting2}
\end{equation}
Comparing (\ref{dtildes1}-\ref{dtildes3}) with (\ref{eq:splitting2}), we
see that the general splitting relations become the splitting conditions in
the effective low energy theory with the matching condition
\begin{equation}
\Lambda_{j,k+1}={\Lambda_j\cdots\Lambda_{k+1}\over\D(Q_{j},\cdots,Q_{k})^{1/2}}.
\label{matching}
\end{equation}
This is the appropriate nonperturbative form of the matching condition in this
case.  Note that this diagonal scale also agrees in form with the diagonal scale
given in the gaugino condensation term of (\ref{eq:noend}).

\section{Higher $n_c$\label{higher}}

It is noteworthy that the discussion in sections \ref{checks} and \ref{ssb}
about
constraints, the splitting relation, and spontaneous symmetry breaking did not
rely in any essential way on the fact that we were specialized to the case
$n_c =2$.  As a matter of fact,
the discussion can be easily generalized to arbitrary $n_c$ by simply
changing $\Lambda^4 \to \Lambda^{2n_c}$. Thus it is reasonable to assume that
these arguments are indeed valid for all $n_c$.\footnote{This has already 
been proven for the 2 site case in \cite{Poppitz:1996wp}}
However, for higher $n_c$, we must solve a more general problem,
because now it is possible for the $SU(n_c)$ gauge groups to break down to
a non-abelian subgroup, $SU(\ell_c)$ for $\ell_c<n_c$. This is a
generalization of the simple 2-site example we worked out in section 3.
Perturbatively, the general classical $D$-flat situation looks
like\footnote{We are
being sloppy here and below with phases. One could (and perhaps should)
keep track of the phases and the
$\theta$ parameters - but we will assume that everything is real and not
worry about these niceties. They are easy to put in correctly because of
holomorphy.}
{\renewcommand{\arraystretch}{2.5}
\begin{equation}
\begin{array}{c}
Q_i =
\pmatrix{
A_{m_c}&0\cr
0&0\cr
}\quad\mbox{for $0\leq i<j$ and $k<i\leq N$}
\\\mbox{and}\quad
Q_i=\pmatrix{
\sqrt{v_i^2I_{m_c}+|A_{m_c}|^2}&0\cr
0&v_iI_{\ell_c}\cr
}\quad\mbox{for $j\leq i\leq k$}
\end{array}
\label{qjs}
\end{equation}
where again $I_{m_c}$ and $I_{\ell_c}$ are $m_c\times m_c$ and
$\ell_c\times\ell_c$ identity
matrices with $m_c+\ell_c=n_c$, and
$A_{m_c}$ is a diagonal $m_c\times m_c$ matrix. In perturbation theory,
this produces an
effective low energy theory like that in fig.~\ref{fig6}, but in which the
sites describe unbroken $SU(\ell_c)$ gauge subgroups, and now all the
fields are suitably modified.

The gauge invariant way of describing this effect makes use of the product,
\begin{equation}
Q_0\cdots Q_N.
\end{equation}
It is the fact that some components of this matrix are large that signals
spontaneous breaking of $SU(n_c)\to SU(\ell_c)$.
To analyze this problem fully, we would find the form of the fields in
terms of the values of the gauge invariant quantities (up to gauge
transformations, of course), and then find appropriate forms for the low
energy fields consistent with the splitting conditions. We will discuss
this further in the next section. Here we will simply introduce the issues
by showing how the splitting conditions work
for one site and $A=aI_{m_c}$. This
is trivial and well understood from the early days of the subject, but it
is worth saying  in our current language.

The perturbative analysis of symmetry breaking in the one site case goes
like this. To
leading order the coupling does not change at the symmetry breaking
threshold, so
\begin{equation}
(\Lambda/a)^{2n_c}=\left (\frac{\mu}{a}\right )^{2n_c}e^{-{8\pi^2}/{g(\mu)^2}}
=(\tilde\Lambda/a)^{2\ell_c}
\label{nctoellc}\end{equation}
or
\begin{equation}
\tilde\Lambda^{2\ell_c}={\Lambda^{2n_c}\over a^{2m_c}}.
\label{nctoellc2}\end{equation}
More generally, the result is
\begin{equation}
\tilde\Lambda^{2\ell_c}={\Lambda^{2n_c}\over \det A_{m_c}^2}.
\label{nctoellc3}\end{equation}
Thus in this case, it is
clear how to make the connection between the high energy theory, with QMMS
\begin{equation}
\d(Q_0Q_1)=\d Q_0 \d Q_1-\Lambda^{2n_c}
\end{equation}
and the low energy theory. The gauge invariant product in the low energy
theory is the projection onto the sector orthogonal to $A_{m_c}$ -
\begin{equation}
Q_0Q_1\to
\pmatrix{
B_{m_c}&0\cr
0&\tilde Q_0\tilde Q_1\cr
}
\quad\mbox{and}\quad
\d(\tilde Q_0\tilde Q_1)={\d(Q_0Q_1)\over\det B_{m_c}}.
\end{equation}
The matrix $B_{m_c}$ is defined nonperturbatively in terms of the
components of the gauge invariant product, $Q_0Q_1$. These components might
be fixed, for example, by Lagrange multiplier terms in the superpotential.
The other gauge invariants would then be related by
\begin{equation}
\d\tilde Q_j=\d Q_j/\sqrt{\det B_{mc}}\quad\mbox{for $j=0$ or $1$}
\end{equation}
Then the QMMS condition in the high energy theory
can be written as
\begin{equation}
\d(\tilde Q_0\tilde Q_1)=\d\tilde Q_0 \d\tilde Q_1-{\Lambda^{2n_c}
\over\det B_{m_c}}
\end{equation}
which is the QMMS condition for the low energy theory with
\begin{equation}
\tilde\Lambda^{2\ell_2}={\Lambda^{2n_c}\over \det B_{m_c}}
\label{nctoellc4}\end{equation}
which is the nonperturbative version of (\ref{nctoellc3}).

\section{Power-law running}
Armed with our new calculational tools, we can attempt to address
a practical example, that of power-law running.  
In the standard analysis \cite{Dienes:1998vh}, power-law running is
not conventional running of the coupling constant at all, but rather just one-loop
quantum corrections to the gauge coupling that are dependent upon the cutoff of the extra-dimensional 
gauge theory.  Thus, it is difficult to know if power-law running is truly UV completion
independent. 

However, if we now consider a theory with a larger $n_c$, we can try to separate the 
issue of power-law running from the physics of the cut-off. Classically, the constraints of $D$-flatness in these theories
are
\begin{equation}
Q_j^\dagger\,Q_j-Q_{j+1}\,Q_{j+1}^\dagger\propto I.
\label{dflat}
\end{equation}
Now, consider
an $SU(n_c)$ chain with equal gauge couplings at the sites and the
following structure of vacuum values for the
$Q$'s consistent with (\ref{dflat}) (ignoring for the moment, the quantum
modifications):
\begin{equation}
Q_0=Q_N=
\pmatrix{
aI_{m_c}&0\cr
0&0\cr
}
\quad\mbox{and}\quad
Q_j=\pmatrix{
\sqrt{v^2+a^2}\,I_{m_c}&0\cr
0&vI_{\ell_c}\cr
}\quad\mbox{for $1\leq j<N-1$}.
\label{qs}
\end{equation}
First suppose that $a=0$. At
energy scales far above $gv$, the theory looks like a four dimensional
theory. But at energies between $gv$ and $gv/N$, we are surrounded by the
KK modes of a five dimensional gauge theory with the fifth dimension
confined to a region of size ${N\over gv}$. Finally, at energies below
$gv/N$, below the mass of the lightest KK mode, our wave-function spreads
over the whole fifth dimension, and we see only the massless unbroken
$SU(n_c)$ gauge theory.  Again, for $a\neq0$, the $SU(n_c)$ gauge group is
broken down to the diagonal $SU(\ell_c)$ and the other gauge multiplets
(all the broken KK modes included) outside this unbroken subgroup get a
contribution to their mass squared.

However, now the question to ask is at what scale does this 5-d $SU(n_c)$ gauge theory 
break down to $SU(\ell_c)$?  To answer this question, we begin by analyzing the 
lowest mass eigenstates of the broken gauge bosons.  Out of the gauge bosons in 
$\frac{SU(n_c)}{SU(\ell_c)}$, the lowest lying state of the $SU(n_c-\ell_c)$ gauge bosons
has a first order contribution $a^2/2N$ whereas the lowest state of the $X,Y$ off diagonal gauge bosons
gets a contribution $\frac{a^4}{4v^2}+\frac{a^2}{N}$.  Out of these scales,
the scales in the $X,Y$ masses are more important from the 5-d point of view.  This is 
because as we go up in energy, only when we encounter the $X,Y$ KK threshholds does the 
beta function change.  Since only they contribute to power law running, the $X,Y$ mass indicates 
when the 5-d theory sees the broken gauge structure.  Out of the two terms in the 
$X, Y$ mass, it turns out that we have to take the first term to dominate over the second
in order to have a continuum limit in which the breaking scale is above the scale of
the extra-dimension (1/R).  
Thus, we will define the breaking scale 
\begin{eqnarray}
v_b \equiv \frac{a^2}{2v}
\end{eqnarray}
and assume 
\begin{equation}
v\gg v_b \gg 1/R \sim v/N.
\end{equation}

However, this theory does not have the continuum limit that we desire.  
The issue is that the continuum limit requires the following $N$ scaling:
\begin{eqnarray}
v \sim N \hspace{0.4in} a \sim \sqrt{N}
\end{eqnarray}
From the form of the 
vevs (\ref{qs}), the 5-d interpretation is of a bulk {\it and} two brane scalar 
fields getting vevs.  The $Q_1, \cdots Q_{N-1}$ vevs correspond to a bulk $SU(n_c)$ adjoint 
getting a vev
\begin{equation}
<\Phi_{bulk}> = \pmatrix{
v_b I_{m_c} & 0 \cr
0 & 0  \cr}
\end{equation}
whereas the $Q_0, Q_N$ vevs correspond to two brane $SU(n_c)$ fundamentals transforming
under a global $SU(n_c)$, one on each brane with vevs
\begin{equation}
<\Phi_{brane}> = \pmatrix{
a I_{m_c} & 0 \cr
0 & 0  \cr}.
\end{equation}
Neglecting the brane vevs, the bulk vev gives a universal contribution to all X,Y KK gauge
bosons, which leads to the standard power-law running.  However, when the bulk and brane
vev contributions become equally important, the KK masses are not universally shifted and
thus the model does not correspond to the standard power-law running setup.  In the 
continuum limit, the brane vevs blow up and give a large threshhold correction to the 
power law running (an explicit calculation confirms this).  
Therefore, in this model, the naive attempt at UV completing the standard power-law 
running setup fails, and thus the techniques formulated in this paper cannot be used
to properly analyze power-law running. 

\section{Conclusion}

In this paper, we have analyzed a set of N-site moose extensions of SUSY
QCD.  In the particular case of $SU(2)$ gauge groups,  we have determined the
superpotentials and the quantum modified moduli constraints of all N-site
chains with or without end links.  By integrating out link fields and setting $\Lambda$'s
to zero, we have discovered
that these results satisfy a restrictive network of interdependent checks.
During the process, we noted that the success of the checks was highly
dependent upon the mathematical properties of the constraints.  

We found that it was particularly useful to recast our constraints in the
form of ``splitting relations'' that are associated with splitting of
determinants of products of link fields when a gauge coupling goes to
zero. The splitting relations are particularly useful when we analyze
regions of moduli space that correspond to a spontaneous breaking of some
of the symmetries of the moose chain. When we combine the splitting
relations with rules for writing the invariants in the low energy theory,
below the symmetry breaking scale, we get nontrivial information about the
physics of the low energy theory. We have shown how to use it to simplify
perturbative analysis and to incorporate nonperturbative results.  
The success and $n_c$ independence of these checks gave compelling evidence
to assume the QMMS constraints are valid for all $n_c$. 

We tried to UV complete a model of power-law running in our setup, but 
the naive attempt did not have the desired continuum limit.  Thus we were not
able to use our calculational tools on this interesting problem.    
As an aside, shortly after this paper was submitted, 
a paper  appeared which analyzed some conditions 
where power-law running corrections can be trusted 
\cite{Hebecker:2002vm}. 

There are directions for further analysis.  Many of our calculations were
specifically for the case of $SU(2)$, so it would be useful to strengthen
the arguments for the form of the constraints for all $n_c$.  Moving on to
other theories, circular moose models have been important to
``deconstructing'' compact extra dimensions.  Some work in this class of
theories has already been done \cite{Csaki:1997zg, Csaki:2001em, Csaki:2001zx}, but there
may still be some
interesting physics to work out in regards to their moduli spaces. In
particular, we believe that the structure of the splitting relations is far
more complicated, and contains far more information than it does for the
linear chains.

\section*{Acknowledgements}

\noindent Nima Arkani-Hamed, Andrew Cohen, and Girma Hailu collaborated with us on some
of the early stages of this work. We thank them for many crucial
discussions and suggestions.  After the first version was submitted to the arXiv, 
Girma Hailu posted two papers containing some overlapping work 
\cite{Hailu:2002bg, Hailu:2002bh}.
We also thank Arthur Hebecker, Erich Poppitz, Yuri Shirman, 
and Alexander Westphal for pointing out issues in the power-law running section.
Some of this work was done at the Aspen Center
for Physics, and HG is very grateful to the staff of the Center and to the
organizers of the ``Advances in Field Theory and Applications to Particle
Physics'' workshop, Andrew Cohen, David Kaplan, Ann Nelson and Matt
Strassler, for the opportunity to work in this stimulating environment.
SC would also like to thank Csaba Cs\'{a}ki for looking over an early draft
of the paper
and for providing helpful comments.
This paper is based upon research that is supported in part by the National
Science Foundation under grant number NSF-PHY/98-02709 and by a National
Science Foundation Graduate Fellowship.

\appendix

\section{Determination of Lower Constraints}
\renewcommand{\arraystretch}{1.2}
This appendix contains another bit of evidence in favor of our assertion
that the lower constraints are local in theory space, satisfying
(\ref{eq:expectedform}).
For the N-site chain, ``integrating in'' has determined the highest constraint
\begin{equation}
\d(Q_0 \cdots Q_N) = \D(Q_0,Q_1, \cdots, Q_N).
\label{eq:largeconstraint}
\end{equation}
Through an argument based on anomaly matching, the constraints on the
smaller determinants can be deduced as well.  For the purposes of this
section, we take general $SU(n_c)$ groups instead of the specific case of
$SU(2)$ to show the generality of the anomaly matching.

First of all, the elementary degrees of freedom in this theory have the
following charges under the non-anomalous global symmetries:

\begin{equation}
\begin{array}[]{c|c c c c}
{\rm Field}     & SU(n_c)_0 & SU(n_c)_{N+1} & U(1) & U(1)_R\\
\hline
   Q_0    &  n_c &      1      &    1   &   -1   \\
   Q_i    &    1    &      1      &    (-1)^i & -1 \\
   Q_N    &    1    & \bar n_c  &    (-1)^N & -1 \\
 \lambda_j  &    1    &      1      &    0        &  1 \\
\end{array}
\end{equation}
where $i = 1, \cdots, N-1$, $j = 1, \cdots, N$,  and the $U(1)_R$ charge is
for the
fermionic component.

This is to be compared with the global charges of the chosen set of gauge
invariant composites:
\begin{equation}
\begin{array}[]{c|c c c c}
{\rm Field}     & SU(n_c)_0 & SU(n_c)_{N+1} & U(1) & U(1)_R\\
\hline
Q_0\cdots Q_N  &  n_c & \bar n_c &   (1+(-1)^N)/2   &   -1   \\
   \d Q_j    &    1    &      1      &    (-1)^j n_c & -1 \\
\end{array}
\end{equation}
where $j = 0, \cdots, N$ and again the $U(1)_R$ charge is for the fermionic
component.

Now, (\ref{eq:largeconstraint}) suggests that there is always a vacuum
where the $U(1)$ is preserved.  Specifically where the vevs of the
determinants are all zero and $<Q_0\cdots Q_N> \; \sim \;
\Lambda_1^2\Lambda_3^2\cdots \Lambda_N^2$ if N is odd and zero if N is
even.  If in addition the anomalies match between composites and elementary
fields, this would convince us that the $U(1)$ preserving vacuum does in
fact exist.

In terms of the elementary fields we get the following global anomalies:
\begin{equation}
\begin{array}[]{ l | c c }
{\rm Global \ Anomaly} & {\rm N \ odd} & {\rm N \ even}\\
\hline
U(1) = U(1)^3  & 0 & n^2\\
U(1)SU(n_c)_0^2 & n/2 & n/2\\
U(1)SU(n_c)_{N+1}^2 & -n/2 & n/2\\
U(1)U(1)_R^2 & 0 & n^2\\
U(1)^2U(1)_R & -(N+1)n^2 & -(N+1)n^2\\
U(1)_R = U(1)_R^3 & -n^2 -N & -n^2 - N\\
U(1)_RSU(n_c)_0^2 = U(1)_RSU(n_c)_{N+1}^2 & -n/2 & -n/2\\
SU(n_c)_0^3 = -SU(n_c)_{N+1}^3 &  n & n \\
\end{array}
\end{equation}

Let's now address the global anomalies on the composite side.  For N even,
let's assume the $U(1)$ preserving vacuum is allowed (i.e. where all vevs
vanish).  In this case, (\ref{eq:largeconstraint}) sets a linear
combination of the $\d Q_j$ with $U(1)$ charge $n_c$ to zero.  Thus, in
matching anomalies we can leave out $\d Q_N$.  The remaining composites $\d
Q_0, \d Q_1, \cdots, \d Q_{N-1}$ and $Q_0\cdots Q_N$ have the exact global
anomalies as in the rightmost column above.

When N is odd, the proposed vacuum that breaks $SU(n_c)_0 \times SU(n_c)_{N+1}
\to SU(n_c)_D$ and preserves $U(1)$ sets $\tr (Q_0Q_1\cdots Q_N)$ to zero.
This leaves an adjoint under the $SU(n_c)_D$.  This adjoint plus all of the
$\d Q_j$ give the following global anomalies:
\begin{equation}
\begin{array}[]{ l |c }
{\rm Composite \ Global \ Anomaly} & {\rm N \ odd}\\
\hline
U(1) = U(1)^3  & 0\\
U(1)SU(n_c)_D^2 & 0\\
U(1)U(1)_R^2 & 0\\
U(1)^2U(1)_R & -(N+1)n^2\\
U(1)_R = U(1)_R^3 & -n^2 -N\\
U(1)_RSU(n_c)_D^2 & -n\\
SU(n_c)_D^3 &  0\\
\end{array}
\end{equation}
These anomalies are also precisely those of the elementary fields.

This analysis has suggested two things: 1) that the full set of QMMS
constraints has an allowed vacuum where the $U(1)$ is preserved and 2) that
the set of composites shown above are truly a complete set (i.e. the other
determinants can be written in terms of them).  That vacuum's existence
along with known limits of the constraints actually completely determines
the QMMS constraints for the smaller determinants.  For instance, consider
the constraint for $\d(Q_0Q_1)$.  The limits $\Lambda_i \to 0$ determine
that $\d(Q_0Q_1) = \d Q_0 \d Q_1 -\Lambda_1^4 + X$ where X has the form
$$ X \sim \sum (\prod_i a_i^{n_i})\Lambda_1^4$$
where $a_i \equiv \Lambda_i^4/(\d Q_{i-1} \d Q_i)$.  However, if the $U(1)$
preserving vacuum truly exists, X must be identically zero.  Thus, the
constraint is just that for the original one site case!

The analysis for the other lower determinants works in the same
fashion.  In each case, the constraint is exactly like the highest
constraint for a lower site model as given in (\ref{eq:expectedform}).
Although it might seem surprising that the non-highest constraints
would take on their na\"{\i}ve form, previous work and observations suggests
that it is to be expected \cite{Poppitz:1996vh}.

\end{document}